\documentclass[conference]{IEEEtran}
\IEEEoverridecommandlockouts
\usepackage{cite}
\usepackage{amsmath,amssymb,amsfonts}
\usepackage{algorithmic}
\usepackage{graphicx}
\usepackage{textcomp}
\usepackage{booktabs}
\usepackage{makecell}
\usepackage{xcolor}
\usepackage{xfrac}
\usepackage{url}
\def\BibTeX{{\rm B\kern-.05em{\sc i\kern-.025em b}\kern-.08em
    T\kern-.1667em\lower.7ex\hbox{E}\kern-.125emX}}
\makeatletter
\newcommand{\linebreakand}{%
\end{@IEEEauthorhalign}
\hfill\mbox{}\par
\mbox{}\hfill\begin{@IEEEauthorhalign}
}
\makeatother
\begin{document}

\title{Low-Density EEG for Seizure Detection: Evaluating CNN-RNN Architectures on a Behind-the-Ear Montage Setup}
% {\footnotesize \textsuperscript{*}Note: Sub-titles are not captured for https://ieeexplore.ieee.org  and

\author{
	\IEEEauthorblockN{Annika Stiehl}
	\IEEEauthorblockA{
	Ansbach UAS\\
	Ansbach, Germany\\
	annika.stiehl@hs-ansbach.de}
	\and
	\IEEEauthorblockN{Patrick Wingert}
	\IEEEauthorblockA{
	Ansbach UAS\\
	Ansbach, Germany\\
	wingert23286@hs-ansbach.de}
	\and
	\IEEEauthorblockN{Nicolas Weeger}
	\IEEEauthorblockA{
	Ansbach UAS\\
	Ansbach, Germany\\
	nicolas.weeger@hs-ansbach.de}

	\linebreakand 
	\IEEEauthorblockN{Nicole Ille}
	\IEEEauthorblockA{
	BESA GmbH\\
	Gräfelfing, Germany\\
	nille@besa.de}
	\and
	\IEEEauthorblockN{Christian Uhl}
	\IEEEauthorblockA{
	Ansbach UAS\\
	Ansbach, Germany\\
	christian.uhl@hs-ansbach.de}
	\and
	\IEEEauthorblockN{Stefan Geißelsöder}
	\IEEEauthorblockA{
	Ansbach UAS\\
	Ansbach, Germany\\
	stefan.geisselsoeder@hs-ansbach.de}
}

\maketitle

\begin{abstract}
Epilepsy affects over 50 million individuals globally, underscoring the need for automated seizure detection systems that can alleviate clinicians’ workload and enhance the accuracy of patient seizure diaries. In wearable EEG applications, however, reliable detection remains challenging due to the limited spatial resolution of low-density electrode configurations, reduced signal-to-noise ratios, and the scarcity of diverse, publicly available training datasets. This study investigates the efficacy of hybrid deep learning architectures for automated seizure detection using a simulated behind-the-ear montage derived from the Temple University Seizure Corpus (TUSZ, v2.0.3). We conduct a systematic comparison of several CNN-RNN models, including LSTM- and GRU-based variants, across multiple EEG montages to evaluate their capacity to compensate for the loss of spatial information inherent to reduced electrode configurations. The proposed CNN-Merged model, which integrates temporal and spectral feature representations, demonstrates superior performance, achieving a ROC AUC of 85.89\,\% and a balanced accuracy of 79.11\,\% on the held-out test set. Furthermore, the model exhibits strong robustness across different reference montages, effectively bridging the performance gap between conventional full-scalp recordings and resource-constrained wearable systems. These findings substantiate the potential of hybrid deep learning models as a promising avenue toward robust, patient-independent seizure detection in low-density EEG applications.
\end{abstract}


\section{Introduction}

Epilepsy remains one of the most prevalent neurological disorders, affecting more than 50 million people worldwide\footnote{\url{www.who.int/news-room/fact-sheets/detail/epilepsy}, accessed: Feb 10, 26}. Accurate and fast seizure detection is essential for effective clinical management. However, manually annotating long-term electroencephalogram (EEG) recordings is time-consuming, error-prone, and requires specialised expertise \cite{SchulzeBonhage2022}. Automated detection systems offer a transformative solution to alleviate the burden on clinicians and improve patient outcomes. Breakthroughs in deep learning, specifically Convolutional Neural Networks (CNNs), Recurrent Neural Networks (RNNs) and Transformer Architectures have demonstrated significant potential in medical EEG signal processing \cite{Prasanna2023, Park2025, Sverrisson2026}. By facilitating end-to-end learning directly from raw data, these models circumvent the need for traditional, manual feature engineering by autonomously extracting representations directly from the signal.

Continuous monitoring of patients via standard full-cap EEG setups is often impractical in daily life due to physical discomfort, social stigma, and restricted mobility \cite{Beniczky2021}. This leads to significant gaps in seizure documentation. Research suggests that 86-94\,\% of absence seizures go unrecorded in patient diaries, underscoring the inherent unreliability of self-reporting \cite{Chatzichristos2022}. Low-density EEG systems, such as behind-the-ear (bte) montages, have emerged as a discreet and unobtrusive alternative for long-term ambulatory monitoring \cite{Vandecasteele2020}. Nevertheless, the transition to a reduced electrode count introduces technical hurdles. Lower signal-to-noise ratios (SNR), increased vulnerability to muscle artifacts, and diminished spatial information make it difficult for traditional machine learning approaches, which rely on handcrafted features, to maintain satisfactory performance \cite{Alarfaj2025,ClaroBhagubai2025}.

Recent studies validate the potential of ear-centered EEG for long-term epilepsy monitoring. These studies demonstrate that automated deep learning models using behind-the-ear signals currently achieve moderate sensitivity (68\,\%) with some false alarms \cite{Alarfaj2025}. However, human expert review of bte EEG data yields high detection rates ($>$85\,\%) comparable to clinical gold standards with minimal false detections \cite{Joyner2024}.
For focal epilepsy, patient-specific algorithms using behind-the-ear EEG achieved a sensitivity of 69.1\,\% with 0.49\,FA/24\,h in a hospital setting, but performance dropped significantly in home environments (23\,\% sensitivity) due to artifacts \cite{Vandecasteele2020, ClaroBhagubai2025}. In contrast, automated detection of absence seizures was more robust, reaching sensitivities of up to 98.3\,\% in the hospital and maintaining approximately 84\,\% sensitivity in home settings when multimodal analysis (including motion data) was used to reduce false alarms \cite{ClaroBhagubai2025, Chatzichristos2022}.

To overcome these limitations, hybrid deep learning architectures have been proposed, merging the spatial feature extraction of CNNs with the temporal dynamic modeling of RNNs, such as Long Short-Term Memory (LSTM) or Gated Recurrent Unit (GRU) networks \cite{Hassan2024, Shanmugam2023}. In these frameworks, CNNs interpret spatial correlations within the limited channels, while RNNs capture the sequential evolution of seizure activity over time. This paper evaluates the efficacy of CNN-RNN hybrid architectures specifically for bte electrodes derived of full-scalp EEG recordings. We investigate whether these advanced models can compensate for the spatial constraints of low-density setups. We also provide a comparative analysis of various EEG montages in the time and frequency domains and of different model architectures, in order to determine their viability for reliable, automated seizure detection.

%\textcolor{red}{\cite{Stiehl2025a}}

\section{Methodology} \label{sec:methods}

\subsection{Dataset} \label{sec:dataset}
% Most of the existing wearable EEG datasets largley relies on private datasets

The development of automated systems for wearable EEG is currently hampered by the scarcity of large-scale, public datasets. While specialized datasets, such as the open-access corpus by Bhagubai et al. \cite{Bhagubai2025}, have recently been made available these are often restricted to specific clinical cohorts, such as patients with focal epilepsy. Since the objective of this work is to achieve a generalized, patient- and seizure-type-independent detection, such specialized datasets do not provide the necessary pathological diversity. 

Consequently, we adopted a simulation-based approach by utilizing the Temple University Hospital (TUH) EEG Seizure Corpus (TUSZ v2.0.3) \cite{Shah2018}. By reducing the spatial resolution of this large-scale, expert-annotated clinical dataset to a behind-the-ear configuration, we evaluated our architectures across a heterogeneous range of seizure types and patient demographics. Our selection was restricted to adult patients ($\ge 18$ years) and recording files with a minimum duration of $3$\,s. We strictly adhered to the official fixed partitioning into training, validation (here \textit{dev}), and test sets (here \textit{eval}) to ensure reproducibility and direct comparability with existing state-of-the-art benchmarks. For training and validation, we utilized the provided seizure and background annotations. Since the test set contains almost no background labels, we treated all unannotated data as background during evaluation. 

% \textcolor{red}{Hier noch der Hinweis, dass in eval fast nur seizure labels sind und keine background labels, deswegen wir alle unannotated data als background behandeln.}

% - TUH EEG Seizure Corpus (TUSZ v2.0.3) [cite] \\
% - Annotadet open access dataset with seizure events by expert neurologists \\
% - Selection: >18 years, Files with at least 3 seconds recording \\
% - Fixed training, validation and test set 

\subsection{Preprocessing} \label{sec:preprocessing}

All EEG signals were resampled to $256$\,Hz and filtered using a 60\,Hz second-order IIR notch filter to suppress powerline noise. A bandpass filter was applied, consisting of a first-order zero-phase Butterworth highpass ($1$\,Hz, $6$\,dB/oct) and a fourth-order zero-phase Butterworth lowpass ($25$\,Hz, $24$\,dB/oct) \cite{Chatzichristos2022}. 

To simulate a bte setup from the TUSZ recordings, we extracted the electrode subset $\{P7, T7, P8, T8\}$, which are the nearest to the ears. From these, we derived two distinct spatial representations: (i) an average montage, where each channel is re-referenced to the mean of the subset, and (ii) a bipolar montage consisting of the pairs adapted from \cite{Vandecasteele2020}: \\$\{P7-T7, P8-T8, P7-P8, T7-T8\}$.

The continuous data was segmented into $2$\,s windows. To ensure data quality, a plausibility check was performed on each segment: windows were excluded if any channel exhibited a standard deviation outside the range of $[0, 5000]$ or if all four channels were identified as inactive (standard deviation $\le 0.01$). Each channel was normalized using a $5\,\%$-$95\,\%$ robust scaler, followed by clipping to the interval $[-0.5, 0.5]$. 
%For specific model variants, the log power spectrum was computed via Fourier Transformation (FT) from the average montage setup for every window.
For specific model variants, the log power spectrum was computed via Fourier Transformation (FT) from the average montage setup for every window. This spectral representation aims to capture frequency-domain characteristics that might be overlooked by time-series inputs alone \cite{Stiehl2025a}.
To improve the training process, we employed data augmentation to the training data by a) flipping the channels along the sagittal midline (left-right swap) and b) using overlapping windows with $0.5\,s$ overlap. 
%\textcolor{red}{We make sure that windows with overlap between Training data and Eval or Test data are not included in the results.} 

% \textcolor{red}{Window level tabelle wie viele seizure und background windows in train, dev, eval? - das wäre nochmal eine Tabelle - da schau ich ob ich den Platz dazu am Ende noch habe}

% - Resampling 256 Hz \\
% - Filtering: 1-25 Hz Bandpass + 60 Hz Notch, butterworth highpass first order 6 dB/oct and butterworth lowpass fourth order 24 dB/oct, second order IIR notch digital filter \\
% - Montage: subset of electrodes (10-20 system) for behind-the-ear montage (P7, T7, P8, T8) out of the full channel setup (no standardize montage setup in TUSZ), Bipolar Montage: P7-T7, P8-T8 and P7-P8, T7-T8\\
% - Segmentation: 2 seconds windows with 50\% overlap \\
% - Normalization: 5\%-95\% standard scaler within each channel over all training data + clipping -0.5 / 0.5\\
% - partly: Fourier Transform (log power spectrum) \\
% - Data Augmentation: Flip along the mid axis (Left-Right)

\subsection{Model Architectural Details} \label{sec:models}

We evaluated three model architectures to investigate the efficacy of low-density bte-EEG monitoring for seizure detection.

\subsubsection{Time-Series Models (CNN-LSTM/GRU)}
The backbone for time-domain signals consists of a 1D-CNN feature extractor coupled with a recurrent stage. The CNN comprises three layers with increasing filter counts $\{32, 64, 128\}$ and decreasing kernel sizes $\{7, 5, 3\}$, each followed by batch normalization, ReLU activation, and dropout ($p\in\{0.3, 0.4\}$). For temporal modeling, we compared a two-layer bidirectional LSTM and a GRU variant (hidden size: $128$). To aggregate temporal features, we concatenated the global mean and maximum across the time dimension:
\begin{equation}
    \mathbf{z}_{time} = [\text{mean}(\mathbf{H}), \text{max}(\mathbf{H})],
\end{equation}
where $\mathbf{H}$ denotes the hidden states of the recurrent layer.
This stage was followed by a fully connected classification head with three layers ($256$, $64$, $2$ neurons), ReLU activations, and dropout ($p=0.4$).

\subsubsection{Spectral Feature Model (CNN-FT)}
A four-layer 1D-CNN is used for frequency-domain analysis. It employs a hierarchical kernel strategy ($k=7, 5, 3, 3$) to capture both broad spectral patterns and fine-grained details. An \textit{adaptive average pooling} layer ensures a fixed-length feature vector $\mathbf{z}_{spec}$, regardless of the input's spectral resolution, followed by a similar classification head of fully connected layers as in the time-series models.

\subsubsection{Hybrid Multi-Input Merged Model (CNN-Merged)}
To leverage both temporal and spectral dynamics, we proposed a hybrid multi-input fusion model. This architecture processes the time-series (average montage) and log-power spectrum in parallel branches. The resulting feature vectors, $\mathbf{z}_{time}$ and $\mathbf{z}_{spec}$, are concatenated and passed to a fully connected classification head. This late-fusion approach allows the model to learn complementary representations from both domains.

\subsection{Training Procedure}
The models were implemented in PyTorch and trained on an NVIDIA A100 GPU for $1000$ epochs. We minimized the cross-entropy loss using the AdamW optimizer (initial $lr=10^{-3}$) with a \textit{ReduceLROnPlateau} scheduler (patience: $50$ epochs) and a batch size of $256$. We addressed the inherent class imbalance by assigning a weight ratio of 1:10 to background and seizure events, respectively, within the loss function. Experiment tracking and hyperparameter optimization were managed via MLflow.

\subsection{Model Selection}
\label{sec:model_selection}

Model selection was performed based on the ROC AUC metric evaluated on the validation set. To ensure the selection of a stable checkpoint and to mitigate the influence of stochastic performance fluctuations during training, we employed a \textit{robust ROC AUC} criterion. Instead of selecting the single highest peak, we monitored performance across a sliding window of $n=3$ consecutive epochs. 

For each epoch, the robust score was defined by penalizing the mean performance with its local volatility:
\begin{equation}
    \text{AUC}_{\text{robust}} = \mu_{\text{AUC}} - 0.5*\sigma_{\text{AUC}},
\end{equation}
where $\mu_{\text{AUC}}$ and $\sigma_{\text{AUC}}$ represent the mean and standard deviation of the ROC AUC scores within the three-epoch window, respectively. The model checkpoint corresponding to the maximum $\text{AUC}_{\text{robust}}$ was selected for final evaluation on the held-out test set.

% - CNN-LSTM architecture \\
% - CNN-GRU architecture \\
% - for power spectrum: CNN only architecture \\
% - for time series data + power spectrum data (8 channels in sum): multi-input model with two branches (CNN-LSTM or CNN-GRU for time series + CNN for power spectrum) + feature concatenation before final dense layers \\ 
% - Hyperparameter tuning by hand (Hyperparameter from preprocessing + model architecture), CNN-LSTM and CNN-GRU similar architectures \\
% - Training: Cross-Entropy Loss, AdamW optimizer, starting learning rate 0.001 with lr scheduler (ReduceLROnPlateau 50 Epochs patients), batch size 256, random seed for reproducibility,  \\
% - experiment tracking: mlflow\\
% - Training on NVIDIA A100 40GB GPU with 32 cores CPU \\
% - Number Epochs: 1000 

\section{Results}

The performance of the proposed architectures was evaluated on both the TUSZ v2.0.3 validation (\textit{dev}) and test (\textit{eval}) sets. To ensure stability, the best model checkpoints were selected using the robust ROC AUC criterion described in sec.~\ref{sec:model_selection}, while classification thresholds for the balanced accuracy were optimized solely on the validation data.

\subsection{Performance on the Validation Set (\textit{dev})}

%\begin{table}[htbp]
%	\centering
%	\caption{Performance comparison of different architectures and montages on the TUSZ v2.0.3 Validation Set (dev) (\%). Results are ranked by ROC AUC.}
%	\label{tab:results_dev}
%	\begin{tabular}{@{}llccc@{}}
%		\toprule
%		\textbf{Architecture} & \textbf{Montage} & \textbf{Trained Epochs} & \textbf{BACC} & \textbf{ROC AUC} \\ \midrule
%		CNN-Merged (Hybrid)   & Multi-Input      & 918                     & \textbf{83.08}& \textbf{90.14}            \\
%		CNN-LSTM              & Average          & 324                     & 81.97         & 90.05            \\
%		CNN-LSTM              & Bipolar          & 612                     & 82.02         & 89.93            \\
%		CNN-GRU               & Average          & 507                     & 78.78         & 89.48            \\
%		CNN-FT                & Spectral         & 908                     & 82.15         & 89.38            \\
%		CNN-GRU               & Bipolar          & 834                     & 79.88         & 88.55            \\ \bottomrule
%	\end{tabular}
%\end{table}

\begin{table}[htbp]
	\centering
	\caption{Performance comparison of different architectures and montages on the validation set in \% (ranked by ROC AUC).}
	\label{tab:results_dev}
	\begin{small}
		\resizebox{\columnwidth}{!}{
		\begin{tabular}{@{}llccccc@{}}
			\toprule
			\textbf{Architecture} & \textbf{Input}   & \makecell[c]{\textbf{Trained} \\ \textbf{Epochs}}	& \textbf{BACC} & \textbf{Sens.} & \textbf{Spec.} & \textbf{ROC AUC} \\ \midrule
			CNN-Merged   		  & Multi-Input    	 & 918                     								& \textbf{83.08}& \textbf{88.06} & 78.61		  & \textbf{90.14}   \\
			CNN-LSTM              & Average          & 324                     								& 82.04         & 78.17			 & \textbf{85.92} & 90.05            \\
			CNN-LSTM              & Bipolar          & 612                     								& 82.17         & 81.36			 & 82.97		  & 89.93            \\
			CNN-GRU               & Average          & 507                     								& 81.77         & 81.03			 & 82.51		  & 89.48            \\
			CNN-FT                & Spectral         & 908                     								& 82.95         & 86.58			 & 79.32		  & 89.38            \\
			CNN-GRU               & Bipolar          & 834                     								& 80.04         & 78.70			 & 82.10		  & 88.55            \\ \bottomrule
		\end{tabular}
	}
	\end{small}
	\smallskip
	\raggedright
\end{table}

%The performance metrics for the validation phase are summarized in Table \ref{tab:results_dev}. The CNN-Merged (Hybrid) architecture utilizing a multi-input domain achieved the highest ROC AUC with $90.14\,\%$, closely followed by the CNN-LSTM model on the average montage with a ROC AUC of $90.05\,\%$. 
%
%Overall, all evaluated configurations demonstrated high competitive performance, with ROC AUC values maintained within a narrow range between $88.55\,\%$ and $90.14\,\%$. This consistency across varying architectures, reference montages, and input domains highlights the robustness of the proposed deep learning frameworks. Notably, the CNN-LSTM model achieved near-peak performance in significantly fewer training epochs ($324$) compared to the hybrid and spectral approaches, suggesting a higher high-level feature extraction efficiency for temporal-sequential modeling in low-density EEG configurations.

The performance metrics for the validation phase are summarized in Table \ref{tab:results_dev}. The CNN-Merged architecture, utilizing the multi-input domain, achieved the highest overall performance with a ROC AUC of 90.14\,\% and the highest balanced accuracy (BACC) of 83.08\,\%. This was closely followed by the CNN-LSTM model on the average montage, which reached a ROC AUC of 90.05\,\% and a BACC of 82.04\,\%.

All the evaluated configurations performed well, with ROC AUC values ranging from 88.55\,\% to 90.14\,\%. A similar consistency is observed in the BACC scores, which ranged from 80.04\,\% to 83.08\,\%. This stability across varying architectures, montages, and input domains highlights the robustness of the proposed deep learning frameworks for seizure detection in low-density EEG configurations.

Notably, the CNN-FT model achieved a high BACC of 82.95\,\%, despite a lower ROC AUC compared to the models trained on time series data. This suggests that spectral features provide a strong basis for class separation when classification thresholds are optimized. Furthermore, the CNN-LSTM model (average) achieved near-peak performance in significantly fewer training epochs (324) compared to the hybrid (918 epochs) and spectral (908 epochs) approaches.

\subsection{Performance on the Test Set (\textit{eval})}

\begin{table}[htbp]
\centering
\caption{Performance Comparison on the test set in~\% (ranked by ROC AUC). Metrics reflect the BACC achieved via optimal thresholding on the validation set.}
\label{tab:eval_results}
\begin{tabular}{@{}llcccc@{}}
\toprule
\textbf{Architecture} & \textbf{Input} & \textbf{BACC} & \textbf{Sens.} & \textbf{Spec.} & \textbf{ROC AUC}      \\ \midrule
CNN-Merged            & Multi-Input      & \textbf{79.11}     & \textbf{81.03} & 77.18			& \textbf{85.89}   \\
CNN-GRU               & Bipolar          & 77.79		      & 78.82          & 76.76          & 84.41			   \\
CNN-FT                & Spectral         & 76.89              & 78.63		   & 75.15          & 84.11            \\
CNN-LSTM              & Average          & 76.76              & 72.01          & \textbf{81.51} & 83.47            \\
CNN-LSTM              & Bipolar          & 76.16              & 73.75          & 78.58          & 83.43            \\
CNN-GRU               & Average          & 74.72              & 69.91          & 79.53			& 81.95  		   \\ \bottomrule
\end{tabular}
\end{table}

Performance on the held-out test set (Table~\ref{tab:eval_results}) was evaluated using the optimal thresholds determined during the validation phase. Although a general performance decrease was observed compared to the validation set. The CNN-Merged model emerged as the most robust architecture, leading with a ROC AUC of $85.89\,\%$ and a BACC of $79.11\,\%$. This confirms that the integration of both temporal and spectral features enhances generalization capabilities when encountering unseen patient data.

%A comparative analysis of the architectures reveals that the hybrid models generally provide a more balanced performance profile. The \textbf{CNN-Merged} variant achieves the highest sensitivity ($81.03\,\%$), effectively identifying seizure activity while maintaining a stable specificity of $77.18\,\%$. In contrast, the RNN-based models—specifically the \textbf{CNN-LSTM} and \textbf{CNN-GRU} using the average montage—exhibit the highest specificity levels, reaching up to $81.51\,\%$. This suggests that temporal modeling on spatial averages is particularly effective at filtering out transient artifacts that might otherwise trigger false positives. 

The CNN-Merged model achieved the highest sensitivity ($81.03\,\%$), demonstrating that combining temporal and spectral domains enhances detection of subtle ictal patterns. While the CNN-LSTM (average) reached peak specificity ($81.51\,\%$) by effectively suppressing false positives through temporal modeling, it exhibited lower sensitivity ($72.01\,\%$). Furthermore, the competitive performance of the spectral-only CNN-FT ($84.11\,\%$ ROC AUC) confirms that frequency-domain information is a powerful discriminator, yet it remains most effective when complemented by raw temporal features. Overall, the multi-input strategy maximizes total yield, whereas RNN-based average-montage models offer an alternative for applications where the reduction of false positives is the primary objective.

\begin{table}[htbp]
\centering
\caption{Performance metrics of the retrained CNN-Merged on the validation and test sets. Results are reported as mean~$\pm$~standard deviation in \%.}
\label{tab:retraining_results}
\begin{tabular}{@{}lcc@{}}
\toprule
\textbf{Dataset} 	 & \textbf{Mean BACC}  & \textbf{Mean ROC AUC}   \\ \midrule
Validation Set       & $82.23 \pm 0.98$    & $89.28 \pm 0.74$        \\
Test Set             & $79.59 \pm 0.85$    & $85.88 \pm 0.98$        \\ \bottomrule
\end{tabular}
\end{table}

To assess the reliability of the findings, the best-performing architecture (CNN-Merged) was retrained with two additional random seeds. As shown in Table~\ref{tab:retraining_results}, minor fluctuations were observed across the different experimental runs. 
As expected due to hyperparameter tuning on the validation set, the mean BACC and ROC AUC decreased by $0.85$\,\% and $0.86$\,\% respectively compared to the reported peak performance. The performance on the test set remained remarkably consistent, with a slight BACC increase of $0.48\,\%$ and a negligible ROC AUC decrease of $0.01\,\%$. 
These results suggest that the proposed hybrid model is highly stable across different weight initializations. Most importantly, the consistency of the evaluation metrics substantiates the model's robust generalization capability and confirms the reported performance.

\subsection{Seizure Type-Specific Performance}

\begin{table}[htbp]
    \centering
    \caption{Seizure type-specific sensitivity comparison on the test set in \%. Results are obtained at a fixed operating point of approx. 75\,\% specificity and are ranked by overall sensitivity.}
    \label{tab:seizure_types}
    \begin{small} % Alternative zu footnotesize, falls das Probleme macht
    \resizebox{\columnwidth}{!}{
    \begin{tabular}{@{}lccc@{}}
        \toprule
        \textbf{Architecture} & \textbf{Overall} & \textbf{Focal} & \textbf{Generalized}  \\ \midrule
        CNN-Merged            & \textbf{82.99}   & \textbf{80.77}& \textbf{87.62}		  \\
        CNN-GRU (Bip.)        & 80.25            & 80.23		 & 80.29          		  \\
        CNN-FT                & 78.74            & 79.57         & 77.02       	  		  \\
        CNN-LSTM (Avg.)       & 77.35            & 78.11         & 75.76          		  \\
        CNN-LSTM (Bip.)       & 77.09            & 78.96         & 73.19          		  \\
        CNN-GRU (Avg.)        & 74.18            & 76.96         & 68.39          		  \\ \bottomrule
    \end{tabular}
    }
    \end{small}
    
    \smallskip
    \raggedright
    \begin{scriptsize}
		Note: Focal seizures include Focal Non-Specific (FNSZ) and Complex Partial (CPSZ). Generalized seizures include Generalized Non-Specific (GNSZ), Tonic (TNSZ), Tonic-Clonic (TCSZ), and Absence (ABSZ) seizures. 
	\end{scriptsize}
\end{table}

%Adapted from \cite{Beniczky2025}.

To evaluate generalization, we analyzed sensitivity across the seizure categories at a fixed specificity of approximately $75\,\%$ (Table~\ref{tab:seizure_types}). The CNN-Merged model achieved the highest overall sensitivity ($82.99\,\%$), leading in both the focal ($80.77\,\%$) and generalized ($87.62\,\%$) seizure categories. These results suggest that integrating temporal and spectral features effectively captures the diverse characteristics of both localized and widespread seizure activity.

%To evaluate the generalization capabilities of the architectures, we analyzed the sensitivity across different seizure types at a fixed operating point of approximately $75\,\%$ specificity (Table~\ref{tab:seizure_types}). The CNN-Merged model achieved the highest overall sensitivity of $82.99\,\%$, driven by its balanced performance across both focal (FNSZ) and generalized (GNSZ) classes. 
%
%A comparison of the architectures reveals distinct sensitivity patterns depending on the signal characteristics of the seizure types. While the CNN-Merged approach demonstrated superior performance for GNSZ ($86.78\,\%$) and CPSZ ($74.82\,\%$), the bipolar RNN-based models (CNN-GRU and CNN-LSTM) reached higher sensitivity for the TNSZ category, in some cases achieving $100\,\%$. This suggests that the recurrent stage, when combined with a bipolar spatial representation, is particularly effective at capturing the specific temporal dynamics of tonic patterns. In contrast, the ABSZ category was consistently detected with near-perfect sensitivity across nearly all model variants. 

\section{Discussion} \label{sec:discussion}

The results demonstrate the effectiveness of various deep learning architectures for seizure detection in simulated bte EEG configurations. The CNN-Merged model emerged as the top-performing architecture in both validation and testing phases. This superiority is attributed to the increased information density provided by the multi-input domain, which simultaneously leverages temporal sequences and frequency components. By integrating both domains, the model captures a more holistic representation of seizure dynamics compared to single-domain approaches. However, training dynamics revealed that while the hybrid model achieves the highest final accuracy, the single-input models, such as the CNN-LSTM, exhibit faster convergence and generalize to the validation set in significantly fewer epochs. This suggests a notable trade-off between the depth of information content and computational training efficiency. Additionally, the CNN-Merged model is bigger in size and requires more computational resources, which may limit its applicability in real-time or resource-constrained environments.

A key finding of this study is the high degree of robustness across different configurations. The minimal performance variance, with ROC AUC values ranging from 88.55\,\% to 90.14\,\% on validation data, indicates that the choice between bipolar and average montages does not significantly impact the discriminative capability of the networks in this low-density setup. Similarly, the competitive results of the purely spectral model (CNN-FT) imply that the networks can extract highly relevant features from the frequency domain alone. Furthermore, the comparison between bidirectional LSTM and GRU units showed no clear favorite; both RNN types achieved comparable results, confirming that the general CNN-RNN structure is the primary driver of performance rather than the specific recurrent unit.

The observed performance gap between the validation and test sets indicates a overfitting on the validation data, a common challenge in EEG analysis. While utilizing the official fixed train/validation/test split ensures direct comparability with results from other research groups, this approach has inherent limitations. To mitigate such overfitting, k-fold cross-validation could be employed to ensure that the model generalizes across a wider variety of patient demographics and recording conditions. However, selecting the "best" model based solely on validation performance may be suboptimal for the test set if significant distribution shifts exist between the two. The discrepancy in noise characteristics and labeling consistency likely contributes to the performance variance and suggests that future work should explore cross-validation techniques to better capture the hidden dynamics of the data and improve the model stability.

%The observed performance gap between the validation and test sets indicates a degree of overfitting on the validation data, a common challenge in deep learning for EEG analysis. While utilizing the official fixed train/validation/test split ensures direct comparability with results from other research groups, this approach has inherent limitations. 
%\textcolor{red}{Wie könnte man das Overfitting vermeiden? Cross-validation}
%However, selecting the "best" model based solely on validation performance may be suboptimal for the test set if significant distribution shifts exist between the two. This is particularly relevant given the labeling discrepancies between the subsets; while the training and validation sets contain explicit background labels, the test set contains almost none. 
%This discrepancy in noise characteristics and labeling consistency likely contributes to the performance variance and suggests that future work should explore more robust cross-validation techniques to better capture the hidden dynamics of the TUSZ corpus.
% Fixed partitions may mask underlying data dynamics or patient-specific variances that are not immediately apparent but influence model generalization. 
% In our evaluation, all unannotated data in the test set was treated as background, whereas the models were trained on specifically labeled background segments. 

% \textcolor{red}{Hier noch ein Satz zu den Ergebnissen mit EKG?}

Most architectures demonstrate consistent performance across both seizure categories, with focal patterns being detected remarkably well. This strong focal performance might suggest that bte electrode positions are well-suited to capture activity near the temporal lobe, possibly explaining the reliable detection of localized seizures.

%
%However, the CNN-Merged model remains the exception, as the integration of spectral features allows its performance on wider, generalized patterns ($87.62\,\%$) to exceed its focal sensitivity. This indicates that while the bte positioning is naturally advantaged for temporal-focal activity, multi-domain fusion is necessary to fully resolve the complex morphologies of generalized seizures.
%Most architectures demonstrate consistent performance across major seizure categories. Due to their distinct morphology and rhythmic, high-amplitude discharges, absence seizures (ABSZ) are the most reliably detected. However, GRU-based models' superior sensitivity to tonic patterns (TNSZ) suggests that these models more effectively capture the specific temporal dynamics of tonic patterns than other models do.

Our results demonstrate high competitiveness when compared to existing literature utilizing full-scalp TUSZ data, despite the significant reduction in spatial resolution. The CNN-Merged model achieves a ROC AUC of $85.89\,\%$, which is directly comparable to the $85.94\,\% \pm 4.45\,\%$ reported by Chybowski et al. \cite{Chybowski2025} using a ConvLSTM approach on full-montage data. Furthermore, our architectures consistently exceed the accuracy levels of $67.68\,\% \pm 13.79\,\%$ reported for EEGWaveNet \cite{Thuwajit2022}, suggesting that merged temporal and spectral preprocessing provides a more robust feature space than temporal-based architectures.

Recent transformer architectures, such as LookAroundNet~\cite{Sverrisson2026}, have advanced full-scalp detection; however, our hybrid CNN-RNN models maintain a narrow performance gap, reinforcing the efficiency of low-density bte setups. It is important to distinguish these results from event-based detection studies, such as Vandecasteele et al. \cite{Vandecasteele2020}, who reported high sensitivities for bte setups but often at the cost of higher false alarm rates in continuous recordings. While our current segment-wise classification provides a balanced trade-off, the transition to true event-based detection remains a distinct challenge. 

These comparisons underscore that high-performance seizure detection does not strictly necessitate full-scalp coverage. Leveraging hybrid architectures and multi-domain inputs allows low-density EEG configurations to achieve the diagnostic depth necessary for reliable monitoring.

%\textcolor{red}{When compared to existing state-of-the-art literature using the TUSZ full-scalp data, our results demonstrate high competitiveness even with only a subset of electrodes.}
%Our CNN-Merged and CNN-LSTM models get comparable performanve against the ConvLSTM approach reported by Chybowski et al. \cite{Chybowski2025}, which achieved a ROC AUC of $0.8594 \pm 0.0445$.
%\textcolor{red}{While Golmohammadi et al. focused on minimizing false alarms with a high specificity of 97.10\,\%, their sensitivity remained lower at 30.83\,\% \textcolor{red}{Das war auch auf dem Level der Anfalls-Detektion so viel ich weiß (das ist ein Unterschied zur Klassifikation von Segmenten)}. - \textbf{eigentlich brauche ich den Golmohammadi nicht!}}
% 
%\textcolor{red}{LookAroundNet \cite{Sverrisson2026}}
%
%\textcolor{red}{Ergebnisse von \cite{Vandecasteele2020} vergleichen}
% 
%In contrast, our models provide a more balanced trade-off, which is essential for applications requiring higher sensitivity without sacrificing significant specificity. Additionally, our architectures consistently exceeded the accuracy levels of $67.68\,\% \pm 13.79\,\%$ reported for EEGWaveNet by Thuwajit et al. \cite{Thuwajit2022}. 
%These comparisons underscore that comparable even better results can be achieved by a low-density EEG setup with merged time and frequency domain preprocessing.
%Also, it shows that CNN-RNN hybrid architectures are well-suited for seizure detection tasks and that only a few electrodes are needed to achieve good performance. 

\section{Conclusion}

This study evaluated various hybrid deep learning architectures for automated seizure detection using simulated low-density behind-the-ear EEG configurations. Our results demonstrate that CNN-RNN frameworks, particularly the CNN-Merged model, with a ROC AUC of $85.89\,\%$ on the held-out test set, can reliably extract discriminative features from sparse electrode setups, achieving performance levels comparable to full-scalp montages. 

Despite these promising results, the current specificity remains insufficient for autonomous daily use, as the frequency of false detections would likely cause significant user burden. To address this, future work must transition from segment-wise classification to robust, event-based detection by incorporating ensemble methods and dedicated post-processing. Furthermore, validating these architectures on real-world wearable datasets \cite{Bhagubai2025} will be essential to assess their performance under authentic noise and artifact conditions. Subsequent efforts will also focus on integrating explainable AI for enhanced interpretability \cite{Zhang2025} and optimizing model architectures for deployment in resource-constrained edge-computing environments.

\section*{Acknowledgment}
This work was supported by Bavarian Ministry of Economic Affairs, Regional Development and Energy (StMWi, Funding number: DIK-2307-0007//DIK0536/01).

We acknowledge the support and HPC resources provided by the Erlangen National HPC Center of the FAU Erlangen-Nürnberg under the BayernKI project \textit{DLonEEGData}. BayernKI funding is provided by Bavarian state authorities. 
\bibliographystyle{IEEEtran}
\bibliography{/media/annika/Daten/Promotion/13_DeepEEG/01_Literatur/deepeeg}

\end{document}